\documentclass[conference]{IEEEtran}
\IEEEoverridecommandlockouts
\usepackage{url}
\usepackage{booktabs}
\usepackage{multirow}
\usepackage{tcolorbox}
\usepackage{tikz}
\usepackage{pgfplotstable}
\usepackage{pgfplots}
\usepackage{balance}

\usepackage{cite}
\usepackage{amsmath,amssymb,amsfonts}
\usepackage{algorithmic}
\usepackage{graphicx}
\usepackage{textcomp}
\usepackage{xcolor}
\def\BibTeX{{\rm B\kern-.05em{\sc i\kern-.025em b}\kern-.08em
    T\kern-.1667em\lower.7ex\hbox{E}\kern-.125emX}}

\definecolor{beaublue}{rgb}{0.74, 0.83, 0.9} 
\definecolor{redbeau}{HTML}{cc5b5b}  
\definecolor{orabeau}{HTML}{ff9f79}

\newcommand{\Rqone}{RQ$_1$: \textit{To what extent do developers make contributions to the ecosystem?}}
\newcommand{\Rqtwo}{RQ$_2$: \textit{What kinds of contributions are made to the ecosystem?}}
\newcommand{\Rqthree}{RQ$_3$: \textit{What are the motivation to contribute to the  ecosystem?}}

\begin{document}

\title{Contributing Back to the Ecosystem: \\ A User Survey of NPM Developers}

\author{\IEEEauthorblockN{Supatsara Wattanakriengkrai, Christoph Treude*, Raula Gaikovina Kula}
\textit{Nara Institute of Science and Technology, Japan}\\
\textit{Singapore Management University, Singapore*}\\
\{wattanakri.supatsara.ws3, raula-k\}@is.naist.jp, ctreude@smu.edu.sg}

\maketitle

\begin{abstract}
    With the rise of the library ecosystem (such as NPM for JavaScript and PyPI for Python), a developer has access to a multitude of library packages that they can adopt as dependencies into their application.
    Prior work has found that these ecosystems form a complex web of dependencies, where sustainability issues of a single library can have widespread network effects.
    Due to the Open Source Software (OSS) nature of third party libraries, there are rising concerns with the sustainability of these libraries.
    In a survey of 49 developers from the NPM ecosystem, we find that developers are more likely to maintain their own packages rather than contribute to the ecosystem. 
    Our results opens up new avenues into tool support and research into how to sustain these ecosystems, especially for developers that depend on these libraries.
We have made available the raw results of the survey at \url{https://tinyurl.com/2p8sdmr3}.
\end{abstract}
\begin{IEEEkeywords}
Libraries, NPM, OSS contributions
\end{IEEEkeywords}

\section{Introduction}
Due to the Open Source Software (OSS) nature of third party libraries, there are rising concerns with the maintenance and sustainability of these libraries.
Since most third-party libraries rely on volunteer (usually unpaid and overworked) contributions, the abandonment of these contributors causes potential risks for the sustainability of these libraries \cite{2017_ruby}.
For example, Valiev et al . \cite{fse2018_sustained} showed that packages with more contributors are less likely to become dormant.
Furthermore, prior work has shown that the sustainability of libraries is not only affected by factors internal 
to the package, but also by the ecosystem as well \cite{Bogart2016breakAPI}, e.g., dependencies \cite{Golzadeh:2019}.
Particularly, threats such as vulnerabilities \cite{chinthanet2021lags, Durumeric2014Heartbleed} and transitive dependency changes \cite{dey2019_promise_mockus} may cause a library to risk becoming
obsolete \cite{kula2018developers}.

The motivation of OSS developers to make contribution to OSS projects is a well-studied research area.
For example, Lee et al. \cite{Lee:icse2017} investigated the motivations of one-time code contributors and found that the developers primarily made contributions to fix bugs that impeded their work.
On the other hand, Bosu et al. \cite{Bosu:emse2019} showed that ideological motives were more frequent among blockchain software developers.
Vadlamani et al. \cite{Vadlamani:icsme2020} revealed that personal drivers are more critical than professional factors in motivating GitHub developers.
Similarly, a recent study by Gerosa et al. \cite{Gerosa2021TheSS} on the motivations of OSS developers provided evidence that intrinsic and internalized motivations, such as learning and intellectual stimulation, are highly relevant to many developers.
Other research focused on the motivation of specific contributor profiles, such as newcomers \cite{hannebauer2017}, quasi-contributors \cite{steinmacher:icse2018}, and students \cite{silva:jss2020}.
A number of research also investigated the correlation between motivation and other attributes such as retention \cite{wu2007empirical}, task effort \cite{ke2010effects}, intention to contribute \cite{wu2011influence}, and participation level \cite{meissonierm2012toward}.
In this work, we would like to specifically highlight the need for developers to contribute to third party libraries, especially in the context of large ecosystems like NPM.
To this extent, we hypothesize that developers do consider making contributions to the ecosystem, especially when they have applications that depend on that library.

\begin{table*}[t]
\centering
\caption{Actual survey questions which are divided into different sections.}
\label{tab:sum-question}
\scalebox{1.13}{
\begin{tabular}{@{}ll@{}}
\toprule
\multirow{4}{*}{\textbf{I. Background}} & Q1: How many years of experience do you have with Node.js? \\
 & Q2: What do you consider is your developer skill level with JavaScript or Node.js? \\
 & Q3: How do you improve your developer skills? \\
 & Q4: What is the skill level of your contributions? \\
  & Q5: Which statement do you agree with \\
   & Q6: How many NPM packages do you maintain/contribute to? \\
 \midrule
\multirow{3}{*}{\textbf{II. Frequency of contributions (RQ1)}}
 & Q7: How often do you maintain/contribute to NPM packages on GitHub per week? \\
 & Q8: Which kinds of packages are you more likely to maintain/contribute to? \\
  & \begin{tabular}[c]{@{}l@{}}Q9: What are some reasons to slow down/stop or increase contributions to\\ \hspace{7mm} own/other packages?\end{tabular} \\
 & \begin{tabular}[c]{@{}l@{}}Q10: Please leave your comments or thoughts especially on any other reasons that\\ \hspace{7mm} we did not include.\end{tabular} \\
 \midrule
 \multirow{3}{*}{\textbf{III. Kind of contributions (RQ2)}} 
 & 
 \begin{tabular}[c]{@{}l@{}}Q11: What kind of contributions do you make in order to maintain/contribute to\\ \hspace{7mm} NPM packages on GitHub?\end{tabular}\\
 &  \begin{tabular}[c]{@{}l@{}}Q12: Please leave your comments or thoughts especially on any kind of contribution that\\ \hspace{7mm} we did not include.\end{tabular} \\ \midrule
\multirow{6}{*}{\textbf{IV. Motivation of contributions (RQ3)}} & Q13: Why do you contribute to other NPM packages, i.e., packages which you do not own? \\
 & Q14: What is the reasoning behind choosing the other package to maintain/contribute to? \\
 & Q15: Which factors do you consider when maintaining/contributing to NPM packages? \\
 & \begin{tabular}[c]{@{}l@{}}Q16: Please leave your comments or thoughts especially on any factors that\\ \hspace{7mm} we did not include.\end{tabular} \\
 \bottomrule
\end{tabular}
}
\end{table*}

\section{Research Method}
\label{study-design}

\subsection{Survey Design}

The first author reviewed the literature \cite{Vadlamani:icsme2020, Subramanian:IEEE2022} of similar types of studies and developed a list of potential survey questions related to each research questions. 
Table \ref{tab:sum-question} shows the list of questions in our survey. Our goal in designing the survey was to keep it as short as possible, while still gathering all of the relevant information.
In Q11, we provide respondents a list of kinds of contributions proposed by Subramanian et al. \cite{Subramanian:IEEE2022} as defined below: 

\begin{itemize}
\item \textit{Documentation:} Changes and additions made to documentation files such as READMEs and/or comments explaining code.

\item \textit{Feature:} Adding new functionality/features to the project.

\item \textit{Bug:} Fixing unexpected behaviour in code.

\item \textit{Refactoring:} Restructuring code to make it more understandable/readable and/or conform to coding standards.

\item \textit{GIT related issues:} Solving merge conflicts, adding elements to .gitignore files and other changes related to GIT.

\item \textit{Test cases:} Adding test cases and/or adding code to facilitate
testing.

\end{itemize}

In Q15, we provide respondents a list of factors that drive developer contributions proposed by Vadlamani and Baysal \cite{Vadlamani:icsme2020} as defined below:

\begin{itemize}
\item \textit{Personal Drivers:} Contributing
for hobby/fun, catering to self needs, and providing
inputs due to altruistic tendencies of developers

\item \textit{Professional Drivers:} Job-related
contributions, including organizational culture, job, expertise and contributions to open source community.

\item \textit{Challenges:} Contributing for challenging their skills.

\end{itemize}

To answer our RQs, we conduct a user study on NPM developers.
We selected the NPM JavaScript ecosystem as it is one of the largest package collections on GitHub, and also has been the focus of recent studies \cite{Kikas:msr2017, Abdalkareem:fse2017, Decan:saner2017}.
The ethics review procedure for this study was approved by the Nara Institute of Science and Technology  Ethics Review Board  (ref. 2021-I-5). 
Similar to Chinthanet et al. \cite{chinthanet2021lags}, we
collected a listing to retrieve a listing of email addresses of NPM developers.
Data collection was performed in November 2020.
To avoid spam mailing, we randomly selected and invited 1,990 developers to be participants.
The survey was sent in March 29, 2021 and was available to the respondents for two weeks, i.e., up until April 12, 2021.
We received 49 responses (a submission rate of 3\%).

\subsection{Data Analysis}

Our data analysis is separated into two parts: close-ended and open-ended question analyses.
For the close-ended questions, we use Plotly library to compute frequency distributions.
For the Q13 open-ended question, we use a card sorting method.
This method is used for finding categories by following the card sorting procedures similar to \cite{spencer2009card}, \cite{zimmermann2016card}.
From this method, we are able to define and build a taxonomy of the developers' motivation to contribute to the ecosystem.
For the Q10, Q12, and Q16 open-ended questions, we do not use the card sorting method as we received only 3-4 responses.

\begin{table}[t]
\centering
\caption{Demographics of the respondents (49 Node.js developers).}
\label{tab:demographic}
\scalebox{0.8}{%
\begin{tabular}{@{}lrr@{}}
\toprule
\multicolumn{3}{c}{\textbf{How many years of experience do you have with Node.js?}} \\ \midrule
\multicolumn{2}{l}{2 to 3 years} & 5 \\
\multicolumn{2}{l}{4 to 5 years} & 15 \\
\multicolumn{2}{l}{6 to 7 years} & 11 \\
\multicolumn{2}{l}{8 to 9 years} & 8 \\
\multicolumn{2}{l}{10 years or more} & 10 \\ \midrule
\multicolumn{3}{c}{\textbf{\begin{tabular}[c]{@{}c@{}}What do you consider is your developer skill level \\ with JavaScript or Node.js?\end{tabular}}} \\ \midrule
\multicolumn{2}{l}{I have general skills} & 12 \\
\multicolumn{2}{l}{I have specialized skills} & 35 \\ 
\multicolumn{2}{l}{Others} & 2 \\ \midrule
\multicolumn{3}{c}{\textbf{How do you improve your developer skills?}} \\ \midrule
\multicolumn{2}{l}{I use contributions in GitHub} & 40 \\
\multicolumn{2}{l}{\begin{tabular}[c]{@{}l@{}} I use other forms \\ (e.g., textbooks, websites, or personal time) \end{tabular}} & 39 \\
\multicolumn{2}{l}{Others} & 7 \\ 
\midrule
\multicolumn{3}{c}{\textbf{What is the skill level of your contributions?}} \\ \midrule
\multicolumn{2}{l}{I only make contributions that match my skill level} & 24 \\
\multicolumn{2}{l}{\begin{tabular}[c]{@{}l@{}}I constantly seek for opportunities where\\ my contributions can improve my skills \end{tabular}} & 23 \\
\multicolumn{2}{l}{Others} & 8 \\ \midrule
\multicolumn{3}{c}{\textbf{Which statement do you agree with}} \\ \midrule
\multicolumn{2}{l}{\begin{tabular}[c]{@{}l@{}} I prefer to maintain my packages before \\ contributing to other packages \end{tabular}} & 31 \\
\multicolumn{2}{l}{\begin{tabular}[c]{@{}l@{}} I prefer to maintain to other packages \\ before contributing to my packages \end{tabular}} & 3 \\
\multicolumn{2}{l}{\begin{tabular}[c]{@{}l@{}} I have equal preference to my packages \\ and other packages \end{tabular} } & 15 \\ \midrule
\multicolumn{3}{c}{\textbf{How many NPM packages do you maintain/contribute to?}} \\ \midrule
 & \textbf{Own packages} & \textbf{Other packages} \\ \midrule
Less than 10 packages & 28 & 41 \\
10 to 20 packages & 10 & 4 \\
21 to 30 packages & 2 & 2 \\
31 to 40 packages & 3 & 1 \\
51 packages or more & 6 & 1 \\ 
\bottomrule
\end{tabular}%
}
\end{table}

Our card sorting method is outlined below.
First, the first two authors were assigned to conduct the card sorting task.
They reviewed all responses and created a ''card" for each of them.
For validation, instead of coding the cards separately in parallel and checking the consistency of the coding results, the two authors coded the cards together \cite{begel2014analyze}, \cite{guzzi2013communication}. Agreement was negotiated along the way \cite{campbell2013coding}, i.e., when the coders had different opinions, we interrupted the process to discuss the discrepancy before continuing on. 
As we had no predefined categories, the coders used an open coding approach \cite{zimmermann2016card} to analyze the cards, whereby new themes emerged during the coding process. 
By reading the cards, we grouped them into meaningful categories; each category had a theme describing the developers' motivation to contribute to the ecosystem.
When no new category was gained from the responses, we decided to code all cards to increase the completeness of the discovered a set of categories. 
Finally, after all the cards were analyzed, a taxonomy of the developers' motivation to contribute to the ecosystem was established.

\section{Findings}
\label{findings}
\subsection{Respondent Demographics}
Our survey received answers from 49 respondents. In the following, we report the demographics of the respondents. There are checkboxes questions in the survey, so not all categories sum to 49. Their demographics are presented in Table \ref{tab:demographic}.
\subsection{\Rqone}

\subsubsection{Frequency of Contributions}

\begin{figure}[t]
    \centerline{\includegraphics[trim=20 35 30 20,clip,width=1\linewidth]{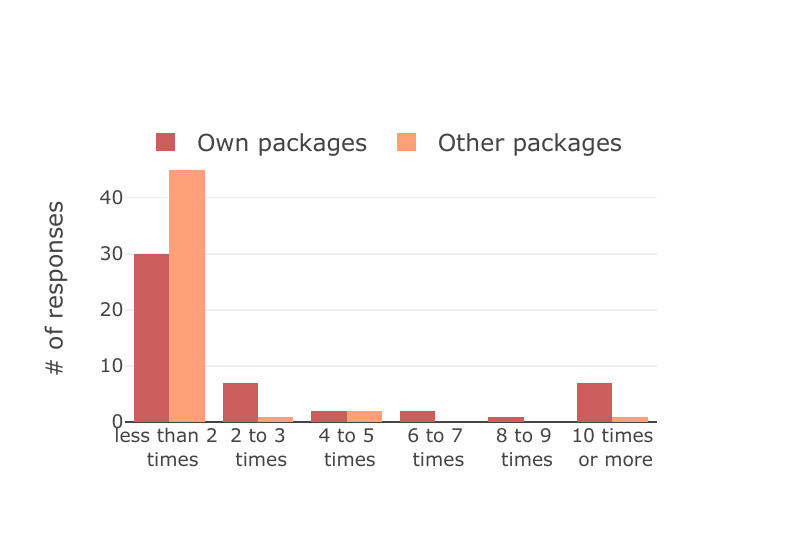}}
    \caption{(Q7) Frequency of contributions between own packages and others.}
    \label{fig:R4}
\end{figure}

From Figure \ref{fig:R4}, we are able to make two observations. First, we find that most of developers maintain their own packages less than 2 times per week and also contribute to other packages less than 2 times per week.
The second observation is that the number of contributions to their own packages per week are higher than contributions to other packages. 
More specifically, 19 respondents said that they maintain their own packages more than 2 times per week, while only 4 respondents said that they contribute to other packages more than 2 times per week. 
These results indicate that NPM developers maintain their own packages more frequently than contribute to the ecosystem.

\begin{figure}[t]
    \centerline{\includegraphics[trim=20 35 20 20,clip,width=1\linewidth]{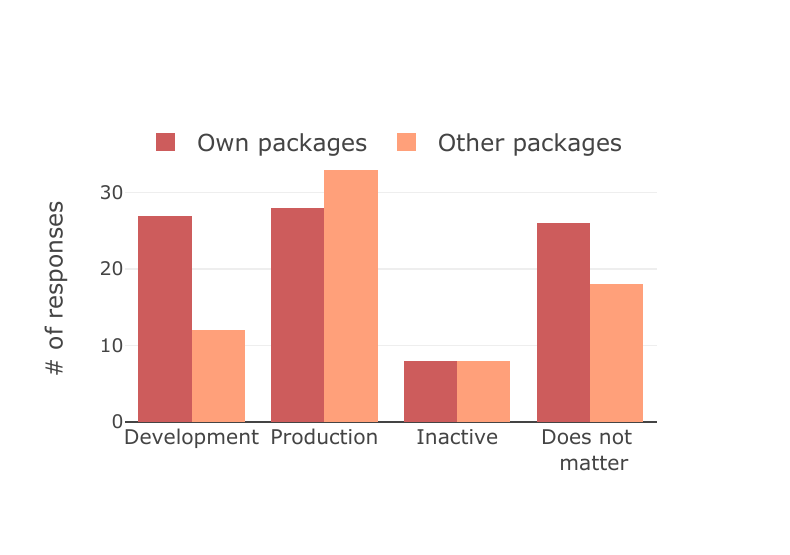}}
    \caption{(Q8) Kinds of packages developers make contributions between own packages and others.}
    \label{fig:R5}
\end{figure}

\subsubsection{Kind of packages developers make contributions}
As shown in Figure \ref{fig:R5}, production is the most common types of packages that developers maintain (28 responses) or contribute to (33 responses), followed by development (27 and 12 responses for their own and other packages, respectively).
These results indicate that NPM developers commonly maintain or contribute stable packages that are already in production.

\subsection{Reasons for stopping or increasing contributions}

\begin{figure}[t]
    \centerline{\includegraphics[trim=20 35 20 20,clip, width=1\linewidth]{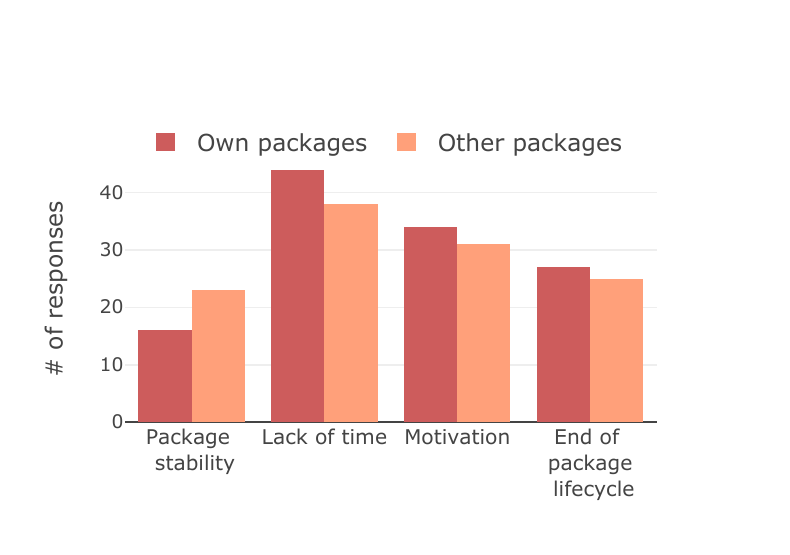}}
    \caption{(Q9) Explanation of why developers stop or increase contributions to own packages and others.}
    \label{fig:R11}
\end{figure}

Figure \ref{fig:R11} shows that lack of time (44 and 38 responses for their and other packages) is the main reason to stop or increase contributions, followed by motivation (34 and 31 responses for their own and other packages) and end of package lifecycle (27 and 25 responses for their own and other packages), respectively.
These findings are consistent between contributions to their own and other packages.
This indicate that developers tend to stop or increase contributions when they are faced with a lack of time.

\subsection{\Rqtwo}

\pgfplotstableread[col sep=comma,]{data.csv}\datatable

\begin{figure}[t]
\centering
\begin{footnotesize}
\begin{tikzpicture}[scale=0.9]
\begin{axis}[
    align=center,
    xbar= 0.0cm,
    bar width=10pt,
    ytick=data,
    yticklabels from table={\datatable}{A},
    xmajorgrids,
    legend style={at={(0.65,0.13)},anchor=south west},
    axis line style={opacity=0},
    major tick style={draw=none},
    point meta=rawx,
    xlabel = \# of responses,
    xmin = 0
    ]
    \addplot [fill=redbeau,draw=none] table [y expr=\coordindex, x=Own packages]{\datatable};
    \addplot [fill=orabeau,draw=none] table [y expr=\coordindex, x=Other packages]{\datatable};
    
    \legend{Own packages, Other packages}
\end{axis}
\end{tikzpicture}
\end{footnotesize}
\caption{(Q11) Kinds of contributions between own packages and others.}
\label{fig:sample_contribution_type}
\end{figure}
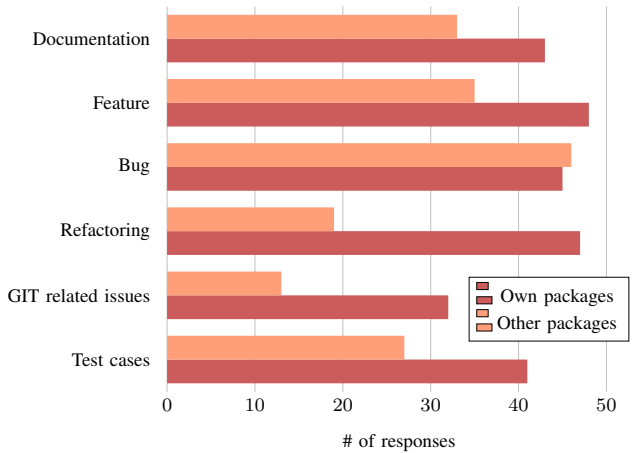

From Figure \ref{fig:sample_contribution_type}, we are able to make two observations.
The first observation is that developers mainly focus on fixing bugs when they contribute to other packages (46 responses).
Compared to fixing their own packages, they tend to spend less efforts on the other kinds of contributions. 
The second most frequent is updating new features (including updating libraries), and then documentation, which is the third most kind of contribution.
Refactoring and GIT related issues were the least contributions to other packages.

Second, we observe that most contributions to their own packages (48 responses) are related to creating a new feature.
These are consistent with the results of Subramanian et al. \cite{Subramanian:IEEE2022}.
The new features may include updating configuration, and third-party libraries.
Not surprising, maintenance activities like refactoring and documentation seem to be high priority contributions for their own packages. 
These results suggest and confirm that developers may focus on making different kinds of contributions depending on whether its their own package or not. 

\subsection{\Rqthree}
To answer this RQ, we summarize results from three perspective: motivations (Q13) and factors of contributions to the ecosystem (Q15 and Q16) and reasons for choosing other packages (Q14).

\subsubsection{Motivations for contributing to the ecosystem}

\begin{table}[t]
\centering
\caption{(Q13) Our coding of motivations for NPM developers in the ecosystem.}
\label{tab:R6-results}
\scalebox{1}{
\begin{tabular}{@{}lr@{}}
\toprule
Reason & \textbf{\# of responses} \\ \midrule
Is a client & 19 \\
Bug Fixing & 15  \\
Ecosystem responsibility& 11 \\
Own motivation& 6  \\ \bottomrule
\end{tabular}
}
\end{table}

\begin{figure}[h]
    \centerline{\includegraphics[trim=20 35 20 20,clip, width=.8\linewidth]{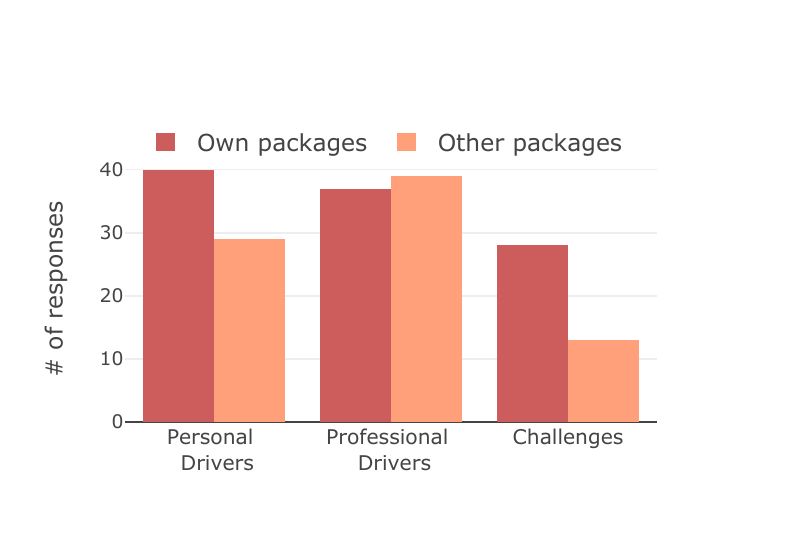}}
    \caption{(Q15) Factors that drive developers to contribute to the ecosystem.}
    \label{fig:R7}
\end{figure}

\begin{figure*}[h]
\centering
\begin{footnotesize}
\begin{tikzpicture}[scale=1]
\begin{axis}[
    align=center,
    y=0.5cm,
    x=0.2cm,
    enlarge y limits={abs=0.25cm},
    symbolic y coords={Others, I am familiar with contributors of that other package, The other package had an open issue that interested me, I wanted to add a new feature to that other package, 
    I depend on that other package, 
    I found a bug in that other package},
    axis line style={opacity=0},
    major tick style={draw=none},
    ytick=data,
    xmajorgrids,
    xlabel = \# of responses,
    xmin = 0,
    point meta=rawx
]
\addplot[xbar,fill=beaublue,draw=none] coordinates {
    (41,I found a bug in that other package)
    (38,I depend on that other package)
    (29,I wanted to add a new feature to that other package)
    (14,The other package had an open issue that interested me)
    (11,I am familiar with contributors of that other package)
    (3,Others)
};
\end{axis}
\end{tikzpicture}
\end{footnotesize}
\caption{(Q14) Reasons why developers choose to contribute to the ecosystem.}
\label{fig:rq1.2}
\end{figure*}
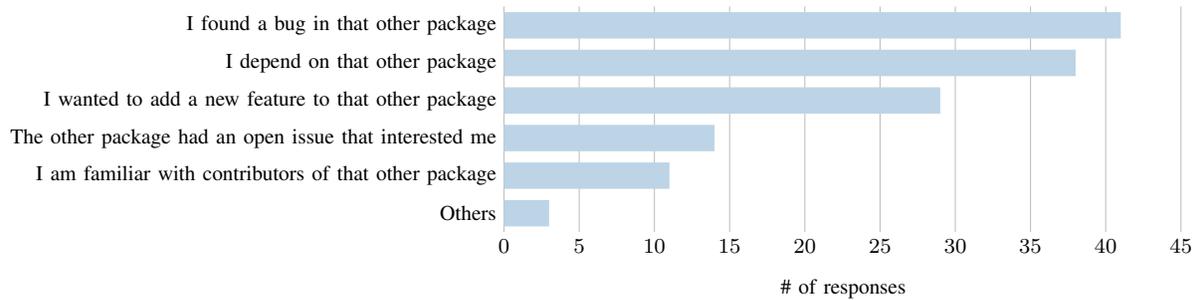

Motivations for contributing to the ecosystem that have emerged during our qualitative analysis are described below.
We find that our manual coding of motivations are similar to the intrinsic and extrinsic motivations found by Gerosa et al.\cite{Gerosa2021TheSS}:

\begin{itemize}
    \item \texttt{Is a client:} Developers contribute to other packages because they want to fix issues on the package that they depend on, e.g., \textit{``I sometimes fix a bug or do an improvement in packages I use''}.
    \item \texttt{Ecosystem responsible:} Developers contribute to other packages because they feel that they are a member of the ecosystem, and therefore would like to solve problems for the greater community, e.g., \textit{``The same reason I make my own packages: to better solve a problem for the entire ecosystem.''}.
    \item \texttt{Bug fixing:} Developers contribute to other packages because they want to fix bugs, acting on their own self-interest, e.g., \textit{``I do my best to fix bugs I come across''}.
    \item \texttt{Own motivation:} Developers contribute to other packages because of other reasons, acting on their own self-interest, e.g., \textit{``not reinvent the wheel again''} 
\end{itemize}

From Table \ref{tab:R6-results}, we are able to make two observations.
First, we find that the main motivation for contributing to other NPM packages is to fix issues related to libraries that they depend on (19 response).
Note that this group includes all kinds of fixes and is not specific to either fixing bugs or adding new features to those packages.
This suggest that developers whose code depends on other packages are more likely to be the ones who contribute to those packages.
Second, when comparing to the related work \cite{Gerosa2021TheSS}, our results did not overlap the more generic motivations of ideology, fun, reputation, learning, career, pay, google summer of code program, and coursework.

\subsection{Factors of contributions}

From Figure \ref{fig:R7}, we are able to make two observations.
First, we observe that personal drivers are the key factors behind developers making contributions to their own packages (40 responses).
They also reserve more challenging tasks for own packages, as oppose to trying to make to other packages.
Second, we find that developers contribute to other packages more due to professional (39 responses) rather than personal drivers.

As part of the  open-ended question on what factors not included in the survey, interestingly, developers report that community is one of the important factors behind their contributions, as one developer mentioned:
\begin{quote}
    \textit{``Building a community is the most difficult part of OSS, supporting an existing projects with an established community is a LOT more valuable than yeeting out something new to flex One's Ego.'' - P19}
\end{quote}

\subsection{Reasons for choosing other packages}

As shown in Figure \ref{fig:rq1.2}, most responses report that they contribute to other packages when they found bugs on those ones (41 responses), while several responses indicate that they prefer to contribute to packages that they depend on (38 responses) or want to add new futures (29 responses).
These results indicate that packages are more likely to receive contributions if developers found bugs in those packages.

\section{Implications}
\label{sec:implications}

\paragraph{Package users and maintainers}
Our results show evidence that package users are indeed considerate of the ecosystem, and look for ways to contribute back to the ecosystem.
As shown in RQ1, developers are aware and make contributions back. 
Furthermore in RQ3, we can see the different reasons for contributing back.
For the package user, we suggest that they do continue to make contributions back to the ecosystem. 
As shown in RQ2, potential contributions can also relate to documentation, new features (updating dependencies), fixing bugs and refactoring.
For package maintainers, our results are welcomed news, as maintainers should look for opportunities for their user base to contribute back.
In this way, maintainers should consider ways to make users aware of existing bugs or opportunities to contribute.

\paragraph{Researchers}
The results of the study show that at developers are aware of the need to contribute back to large ecosystem such as NPM.
As shown in the results, there is more motivation especially if it is to fix a bug that interrupts their work.
Hence, we suggest that tool support is need to help these package users identify situations and opportunities to give back to the ecosystem.
For researchers, this study calls for more research into how the different relationships between developers of a package may assist it with its sustainability. 
From these results, our plan is to conduct a quantitative study to explore the extent to which contributions to the ecosystem can affect the package usage and sustainability, especially to fix bugs, and update the documentation in the library.
Finally, for tool builders, we believe that our research opens up new research into tool support for users to identify potential contributions, similar to the good first issue for newcomers.

\section{Threats to Validity}
\label{validity}
The threat to construct validity is related to the risk of survey respondents misunderstanding the survey questions.
To mitigate this threat, the first two authors discussed the questions with each other, made necessary clarifications of the wording of questions and answers, and provided the literature \cite{Vadlamani:icsme2020, Subramanian:IEEE2022} in the survey.
We also sent the pilot survey to confirm the respondent understanding in the survey questions.

The first threat is related to our taxonomy classification being prone to human-error.
Classified developers' motivation may be miscoded due to the subjective nature of our coding approach.
To mitigate this threat, we made sure that the first two authors who coded the cards had to come to a consensus, and all disputes needed to be carefully considered along with the different opinions. Hence, there were many times where we interrupted the process to discuss the discrepancy before continuing on.
As the submission rate is not high (3\%), our results could suffer from a potential non-response bias, i.e., the responses of the developers who chose to participate may be different from who did not.
However, the 49 responses that we analyzed provide a rich source of data to reveal the insights described in this paper.

\section{Conclusion and Future Directions}
\label{conclusions}

Most developers are more keen to contribute to packages on which they rely, and are mainly driven by fixing a bug in that package.
Package maintainers can leverage our results to revisit current strategies to attract contributions from package users.
Future work includes research into encouraging these potential contributions and understanding the effects of different developer contributions on the package sustainability.
We envision that this work can spark awareness
to support third-party libraries in the ecosystem.

\section*{Acknowledgement}
This work is supported by the Japanese Society for the Promotion of Science (JSPS) KAKENHI Grant Number JP20H05706.

\balance
\bibliographystyle{ieeetr}
\bibliography{reference}

\begin{thebibliography}{10}

\bibitem{2017_ruby}
E.~Constantinou and T.~Mens, ``Socio-technical evolution of the ruby ecosystem in github,'' in {\em 2017 IEEE 24th International Conference on Software Analysis, Evolution and Reengineering (SANER)}, pp.~34--44, 2017.

\bibitem{fse2018_sustained}
M.~Valiev, B.~Vasilescu, and J.~Herbsleb, ``Ecosystem-level determinants of sustained activity in open-source projects: A case study of the pypi ecosystem,'' in {\em Proceedings of the 2018 26th ACM Joint Meeting on European Software Engineering Conference and Symposium on the Foundations of Software Engineering}, ESEC/FSE 2018, p.~644–655, 2018.

\bibitem{Bogart2016breakAPI}
C.~Bogart, C.~K\"{a}stner, J.~Herbsleb, and F.~Thung, ``How to break an api: Cost negotiation and community values in three software ecosystems,'' in {\em Proceedings of the 2016 24th ACM SIGSOFT International Symposium on Foundations of Software Engineering}, FSE 2016, (New York, NY, USA), p.~109–120, Association for Computing Machinery, 2016.

\bibitem{Golzadeh:2019}
M.~Golzadeh, ``Analysing socio-technical congruence in the package dependency network of cargo,'' in {\em FSE2019SRCAffiliation: Fonds de la Recherche Scientifique}, 08 2019.

\bibitem{chinthanet2021lags}
B.~Chinthanet, R.~G. Kula, S.~McIntosh, T.~Ishio, A.~Ihara, and K.~Matsumoto, ``Lags in the release, adoption, and propagation of npm vulnerability fixes,'' {\em Empirical Software Engineering}, vol.~26, no.~3, pp.~1--28, 2021.

\bibitem{Durumeric2014Heartbleed}
Z.~Durumeric, F.~Li, J.~Kasten, J.~Amann, J.~Beekman, M.~Payer, N.~Weaver, D.~Adrian, V.~Paxson, M.~Bailey, and J.~A. Halderman, ``The matter of heartbleed,'' in {\em Proceedings of the 2014 Conference on Internet Measurement Conference}, IMC '14, (New York, NY, USA), p.~475–488, Association for Computing Machinery, 2014.

\bibitem{dey2019_promise_mockus}
T.~Dey, Y.~Ma, and A.~Mockus, ``Patterns of effort contribution and demand and user classification based on participation patterns in npm ecosystem,'' PROMISE'19, p.~36–45, 2019.

\bibitem{kula2018developers}
R.~G. Kula, D.~M. German, A.~Ouni, T.~Ishio, and K.~Inoue, ``Do developers update their library dependencies?,'' {\em Empirical Software Engineering}, vol.~23, no.~1, pp.~384--417, 2018.

\bibitem{Lee:icse2017}
A.~Lee, J.~C. Carver, and A.~Bosu, ``Understanding the impressions, motivations, and barriers of one time code contributors to floss projects: A survey,'' in {\em 2017 IEEE/ACM 39th International Conference on Software Engineering (ICSE)}, pp.~187--197, 2017.

\bibitem{Bosu:emse2019}
A.~Bosu, A.~Iqbal, R.~Shahriyar, and P.~Chakraborty, ``Understanding the motivations, challenges and needs of blockchain software developers: A survey,'' {\em Empirical Software Engineering}, vol.~24, no.~4, pp.~2636--2673, 2019.

\bibitem{Vadlamani:icsme2020}
S.~L. Vadlamani and O.~Baysal, ``Studying software developer expertise and contributions in stack overflow and github,'' in {\em 2020 IEEE International Conference on Software Maintenance and Evolution (ICSME)}, pp.~312--323, 2020.

\bibitem{Gerosa2021TheSS}
M.~A. Gerosa, I.~S. Wiese, B.~Trinkenreich, G.~J.~P. Link, G.~Robles, C.~Treude, I.~Steinmacher, and A.~Sarma, ``The shifting sands of motivation: Revisiting what drives contributors in open source,'' {\em 2021 IEEE/ACM 43rd International Conference on Software Engineering (ICSE)}, pp.~1046--1058, 2021.

\bibitem{hannebauer2017}
C.~Hannebauer and V.~Gruhn, ``On the relationship between newcomer motivations and contribution barriers in open source projects,'' in {\em Proceedings of the 13th International Symposium on Open Collaboration}, pp.~1--10, 2017.

\bibitem{steinmacher:icse2018}
I.~Steinmacher, G.~Pinto, I.~S. Wiese, and M.~A. Gerosa, ``Almost there: A study on quasi-contributors in open-source software projects,'' in {\em 2018 IEEE/ACM 40th International Conference on Software Engineering (ICSE)}, pp.~256--266, IEEE, 2018.

\bibitem{silva:jss2020}
J.~O. Silva, I.~Wiese, D.~M. German, C.~Treude, M.~A. Gerosa, and I.~Steinmacher, ``Google summer of code: Student motivations and contributions,'' {\em Journal of Systems and Software}, vol.~162, p.~110487, 2020.

\bibitem{wu2007empirical}
C.-G. Wu, J.~H. Gerlach, and C.~E. Young, ``An empirical analysis of open source software developers’ motivations and continuance intentions,'' {\em Information \& Management}, vol.~44, no.~3, pp.~253--262, 2007.

\bibitem{ke2010effects}
W.~Ke and P.~Zhang, ``The effects of extrinsic motivations and satisfaction in open source software development,'' {\em Journal of the Association for Information Systems}, vol.~11, no.~12, p.~5, 2010.

\bibitem{wu2011influence}
C.-G. Wu, J.~H. Gerlach, and C.~E. Young, ``The influence of open source software volunteer developers’ motivations and attitudes on intention to contribute,'' {\em International Journal of Open Source Software and Processes (IJOSSP)}, vol.~3, no.~4, pp.~24--48, 2011.

\bibitem{meissonierm2012toward}
R.~Meissonierm, I.~Bourdon, S.~Amabile, and S.~Boudrandi, ``Toward an enacted approach to understanding oss developer’s motivations,'' {\em International Journal of Technology and Human Interaction (IJTHI)}, vol.~8, no.~1, pp.~38--54, 2012.

\bibitem{Subramanian:IEEE2022}
V.~N. Subramanian, I.~Rehman, M.~Nagappan, and R.~Kula, ``Analyzing first contributions on github: What do newcomers do?,'' {\em IEEE Software}, vol.~39, pp.~93--101, jan 2022.

\bibitem{Kikas:msr2017}
R.~Kikas, G.~Gousios, M.~Dumas, and D.~Pfahl, ``Structure and evolution of package dependency networks,'' MSR '17, p.~102–112, 2017.

\bibitem{Abdalkareem:fse2017}
R.~Abdalkareem, O.~Nourry, S.~Wehaibi, S.~Mujahid, and E.~Shihab, ``Why do developers use trivial packages? an empirical case study on npm,'' in {\em Proceedings of the 2017 11th Joint Meeting on Foundations of Software Engineering}, ESEC/FSE 2017, p.~385–395, ACM, 2017.

\bibitem{Decan:saner2017}
A.~Decan, T.~Mens, and M.~Claes, ``An empirical comparison of dependency issues in oss packaging ecosystems,'' in {\em 2017 IEEE 24th International Conference on Software Analysis, Evolution and Reengineering (SANER)}, pp.~2--12, 2017.

\bibitem{spencer2009card}
D.~Spencer, {\em Card sorting: Designing usable categories}.
\newblock Rosenfeld Media, 2009.

\bibitem{zimmermann2016card}
T.~Zimmermann, ``Card-sorting: From text to themes,'' in {\em Perspectives on data science for software engineering}, pp.~137--141, Elsevier, 2016.

\bibitem{begel2014analyze}
A.~Begel and T.~Zimmermann, ``Analyze this! 145 questions for data scientists in software engineering,'' in {\em Proceedings of the 36th International Conference on Software Engineering}, pp.~12--23, 2014.

\bibitem{guzzi2013communication}
A.~Guzzi, A.~Bacchelli, M.~Lanza, M.~Pinzger, and A.~Van~Deursen, ``Communication in open source software development mailing lists,'' in {\em 2013 10th Working Conference on Mining Software Repositories (MSR)}, pp.~277--286, IEEE, 2013.

\bibitem{campbell2013coding}
J.~L. Campbell, C.~Quincy, J.~Osserman, and O.~K. Pedersen, ``Coding in-depth semistructured interviews: Problems of unitization and intercoder reliability and agreement,'' {\em Sociological methods \& research}, vol.~42, no.~3, pp.~294--320, 2013.

\end{thebibliography}

\end{document}